\documentstyle[preprint,eqsecnum,aps]{revtex}

\begin{document}
\draft
\title{Modelling Dark Energy with Quintessence and a Cosmological Constant}
\date{\today}
\author{Rolando Cardenas\thanks{rcardenas@mfc.uclv.edu.cu}, Tame Gonzalez\thanks{tame@mfc.uclv.edu.cu}, Osmel Martin\thanks{osmel@mfc.uclv.edu.cu}and Israel Quiros\thanks{israel@mfc.uclv.edu.cu}}
\address{Departamento de Fisica. Universidad Central de Las Villas. Santa Clara. CP: 54830 Villa Clara. Cuba}
\maketitle

\begin{abstract}
In this talk we present a model of the universe in which dark energy
is modelled explicitely with both a dynamical quintessence field and a
cosmological constant. Our results confirm the possibility of a
collapsing universe (for a given region of the parameter space), which
is advantageous  for an adequate formulation  of both perturbative quantum
field and string theories. We have also reproduced the measurements of modulus distance from supernovae with good accuracy.  
\end{abstract}

\section{Introduction}
From 1998 to date several important discoveries in the astrophysical
sciences have being made, which have given rise to the so called New
Cosmology \cite{turner1,turner2}. Amongst its more important facts we
may cite: 

- Flat, critical density accelerating universe

- Early period of rapid expansion (inflation)

- Density inhomogeneities produced from quantum fluctuations during
inflation

- Composition: 2/3 dark energy; 1/3 dark matter; 1/200 brigh stars

- Matter content: $(29\pm 4)\%$ nonbarionic dark matter; $(4\pm 1)\%$
			  baryons, $(0.1-5)\%$ neutrinos

- $T_0=2.275\pm 0.001 K$

- $t_0=14\pm 1 Gyr$

- $H_0=72\pm 7 km.s^{-1} Mpc^{-1}$ 

It is a fact that the standard LCDM model, though rather simple from the theoretical
point of view, can accomodate most of today's astrophysical
data. However, it still has several open questions (see, for instance,
\cite{lahav,pr}), one of them being: could other (yet unknown)
models fit the data equally well?

Alternative models should obviously consider the main components of
the universe. There is much concern about unveiling the dark part of
our universe, which implies that we don't lack candidates. So, for
dark matter we have neutrinos, axions, neutralinos; for dark energy:
the cosmological constant, scalar fields (for example, quintessence),
cosmic field defects, etc. So far, most models of dark energy have a
rather phenomenological character, though a few proposals concerning
the possible role of the dark energy field in the context of
fundamental physics have appeared \cite{fhsw,wetterich,abrs,gasperini,gpv,pietroni}. 

The recent dark energy scalar field research has several interesting
features (see \cite{pr} for an extense review). Many models have
attractor or tracker behaviour, allowing, for a wide range of initial  conditions, a subdominant field energy
density at high redshifts (radiation and matter  dominated eras).

In the simplest versions, scalar fields models of dark energy have a
scalar field kinetic term, and the scalar field is coupled only to
itself and gravity. So, the scalar field part of the model is fully
characterized by the scalar field potential, with some broad
constraints on the initial conditions for the field, if the attractor
behaviour is realized.

Many different potentials have being used (see reviews
\cite{pr,sahni}):
\begin{table*}[tbh!]
\begin{center}
\begin{tabular}{lll}
\hline
Quintessence Potential & Reference\\\hline
& \\
$V_0\exp{(-\lambda\phi)}$ & Ratra \& Peebles (1988), Wetterich (1988), \\
& Ferreira \& Joyce (1998)\\
& \\
$m^2\phi^2, \lambda\phi^4$ &  Frieman et al (1995)\\
& \\
$V_0/\phi^\alpha, \alpha > 0$ &  Ratra \& Peebles (1988) \\
& \\
$V_0\exp{(\lambda\phi^2)}/\phi^\alpha, \alpha > 0$ & Brax \& Martin (1999,2000)\\
& \\
$V_0(\cosh{\lambda\phi} - 1)^p$, & Sahni \& Wang (2000)\\
& \\
$V_0 \sinh^{-\alpha}{(\lambda\phi)}$, & Sahni \& Starobinsky (2000),
Ure\~{n}a-L\'{o}pez \& Matos (2000)\\
& \\
$V_0(e^{\alpha\kappa\phi} + e^{\beta\kappa\phi})$ & Barreiro, Copeland \& Nunes (
2000)\\
& \\
$V_0(\exp{M_p/\phi} - 1)$, & Zlatev, Wang \& Steinhardt (1999)\\
& \\
$V_0[(\phi - B)^\alpha + A]e^{-\lambda\phi}$, & Albrecht \& Skordis (2000)\\

& \\
\hline
\end{tabular}
\caption{}
\end{center}
\end{table*}
 
In this talk I want to call the attention to
exponential potentials, which have being often discarded on fine tunig
arguments or (the simplest exponential) because they can not produce
the wanted transition from subdominant to dominant energy density
(\cite{pr}). However, as shown in \cite{rs,rse}, they have proved useful in
describing several features in the history of the universe, from
radiation decoupling to nowadays. Also, several authors have recently pointed out that the degree of
fine tuning needed in these scenarios is no more than in others usually
accepted \cite{cline,kl,rs}.

Especially interesting results are obtained if we model dark energy
using both a scalar field and a cosmological constant.

The cosmological constant can be incorporated into the quintessence
potential as a constant which shifts the potential value, especially, the
value of the minimum of the potential, where the quintessence field rolls
towards. Conversely, the height of the minimum of the potential can also be
regarded as a part of the cosmological constant. Usually, for separating
them, the possible nonzero height of the minimum of the potential is
incorporated into the cosmological constant and then set to be zero. The
cosmological constant can be provided by various kinds of matter, such as
the vacuum energy of quantum fields and the potential energy of classical
fields and may also be originated in the intrinsic geometry. So far there is
no sufficient reason to set the cosmological constant (or the height of the
minimum of the quintessence potential) to be zero, especially when the
ultimate fate of our universe is more sensitive to the presence of the
cosmological constant (or the nonzero height of the minimum of the
quintessence potential) than any other matter content, even though the
cosmological constant may be extremely tiny and undetectable at all in
present time (\cite{hwang}. In
particular, some mechanisms to generate a negative cosmological constant
have been pointed out, in the context of spontaneous symmetry breaking
\cite{ss,gh}.
 
\section{The model}
 We consider a model consisting of a three-component cosmological
fluid: matter, scalar field (quintessence with an exponential potential) and a negative cosmological constant. We point out that we model dark energy with both the quintessence field and the negative cosmological constant, resulting possitive our effective cosmological constant, in agreement with experimetal data \cite{efsetal}. ''Matter'' means barionic + cold dark matter, with no
pressure, and the scalar field is minimally coupled and noninteracting with
matter, so the action is: 

\begin{equation}
S=\int d^4 x\;\sqrt{-g}\{\frac{c^2}{16\pi G}(R-2\Lambda)+{\cal L}_\phi+{\cal %
L}_{m}\},
\end{equation}
where $\Lambda$ is the cosmological constant, ${\cal L}_{m}$ is the
Lagrangian for the matter degrees of freedom and the Lagrangian for the
quintessense field is given by

\begin{equation}
{\cal L}_{\phi}=-\frac{1}{2}\phi_{,n} \phi^{,n}-V(\phi).
\end{equation}
This model cannot be used from the very beginning of the universe,
but only since decoupling of radiation and dust. Thus, we don't take into
account inflation, creation of matter, nucleosynthesis, etc.
We apply the same technique of adimensional variables we used in
\cite{cmq} (this allows to determine the integration constants without
additional assumptions). We use the dimensionless time variable $\tau =H_{0}t$, where $t$ is
the cosmological time and $H_{0}$ is the present value of the Hubble
parameter. In this case $a(\tau )=\frac{a(t)}{a(0)}$ is the scale factor. Then we have that, at present $(\tau =0)$

\begin{eqnarray}
a(0)&=&1,  \nonumber \\
\dot{a}(0)&=&1,  \nonumber \\
H(0)&=&1,
\end{eqnarray}

\bigskip Considering a homogeneous and isotropic universe, and using
the experimental fact of a spatially flat universe \cite{bernardis},
the field equations derivable from (2.1) are

\begin{equation}
(\frac{\dot{a}}{a})^2=\frac{2}{9}\sigma^2 \{ \frac{\bar D}{a^3}+\frac{1}{2} 
\dot{\phi}^2 + \bar V(\phi) \ + \frac{3}{2}\frac{\bar{\Lambda}}{\sigma^2}\},
\end{equation}

\begin{equation}
2\frac{\ddot{a}}{a}+(\frac{\dot{a}}{a})^2=-\frac{2}{3}\sigma^2 \{ \frac{1}{2}%
\dot{\phi}^2-\bar V(\phi) \ - \frac{3}{2}\frac{\bar{\Lambda}}{\sigma^2}\}\},
\end{equation}
and

\begin{equation}
\ddot{\phi}+3\frac{\dot{a}}{a}\dot{\phi}+\bar{V}^{\prime}(\phi)=0,
\end{equation}
where the dot means derivative in respet to $\tau$ and,

\begin{eqnarray}
\bar{V}(\phi)&=&\bar{B}^2 e^{-\sigma\phi},
\end{eqnarray}

\begin{eqnarray}
\bar{X} &=&\frac{X }{H_{0}^{2}},  \nonumber \\ (except for 
\bar{D} &=&\frac{D}{a_{0}^{3}H_{0}^{2}}=\frac{\rho _{m_{0}}}{H_{0}^{2}}),
\end{eqnarray}
with $\rho _{m_{0}}$ - the present density of matter, $\sigma ^{2}=\frac{12\pi G%
}{c^{2}}$ and $B^{2}$ - a generic constant.

Applying the Noether Symmetry Aproach \cite{rmrs,r,crrs,rr}, it can be shown that the new variables we should introduce to simplify the
field equations are the same used in \cite{rs}: 

\begin{equation}
a^3=u v,
\end{equation}
and

\begin{equation}
\phi=-\frac{1}{\sigma}\ln(\frac{u}{v}).
\end{equation}

In these variables the field Eqs. (2.4-2.6) may be written as the following pair of equations

\begin{equation}
\frac{\ddot{u}}{u}+\frac{\ddot{v}}{v}=\bar{B}^2\sigma^2\frac{u}{v}-\sigma^2%
\bar{V}_{0},
\end{equation}
and

\begin{equation}
\frac{\ddot{u}}{u}-\frac{\ddot{v}}{v}=-\sigma ^{2}\bar{B}^{2}\frac{u}{v},
\end{equation}
where:

\begin{equation}
\bar{V}_{0}=-\frac{3}{2}\frac{\bar\Lambda}{\sigma^2}.
\end{equation}

Combining of Eqs. (2.11) and (2.12)yields

\begin{equation}
\ddot{u}=-\frac{\sigma^2\bar{V}_{0}}{2}u,
\end{equation}

and

\begin{equation}
\ddot{v}=-\frac{\sigma^2\bar{\vee}_{0}}{2}v+\sigma^2\bar{B}^2u.
\end{equation}

The solutions of the equations (2.14) and (2.15) are found to be

\begin{equation}
u(\tau)=u_{1}\sin(\sigma\sqrt{\frac{\bar{V}_0}{2}}\tau)+u_{2}\cos(\sigma%
\sqrt{\frac{\bar{V}_0}{2}}\tau),
\end{equation}
and

\begin{eqnarray}
v(\tau ) &=&\{v_{2}+\frac{\bar{B}^{2}}{2\bar{V}_{0}}u_{2}-\frac{\sigma \bar{B%
}^{2}}{\sqrt{2\bar{V}_{0}}}u_{1}\tau \}\cos (\sigma \sqrt{\frac{\bar{V}_{0}}{%
2}}\tau )+  \nonumber \\
&+&\{v_{1}+\frac{\bar{B}^{2}}{2\bar{V}_{0}}u_{1}+\frac{\sigma \bar{B}^{2}}{%
\sqrt{2\bar{V}_{0}}}u_{2}\tau \}\sin (\sigma \sqrt{\frac{\bar{V}_{0}}{2}}%
\tau ),
\end{eqnarray}
where $u_{1},u_{2},v_{1}$ and $v_{2}$ are the integration constants. 

In finding the integration constants we use the equations (2.3) and field equations evaluated at $\tau =0$.
Finally, using $\Omega_{m_0}+\Omega_{Q_0}+\Omega_{%
\Lambda}=1 $ and the ansatz

\begin{equation}
\bar B^2=n\;\bar V_0,
\end{equation}
where $n$ is a positive real number, then the above integration constants
can be written in the following way:

\begin{equation}
u_2^{(\pm)}=\pm\sqrt{\frac{2-q_0-1.5\Omega_{m_0}-3\Omega_\Lambda}{%
-3n\Omega_\Lambda}},
\end{equation}

\begin{equation}
v_2^{(\pm)}=\frac{1-\frac{n}{2}\;u_2^2}{u_2^{(\pm)}},
\end{equation}

\begin{equation}
u_{1\;[\pm]}^{(\pm)}=\frac{\{\sqrt{3}-[\pm]\sqrt{1+q_0-1.5\Omega{m_0}}\}}{%
\sqrt{-3\;\Omega_\Lambda}}u_2^{(\pm)},
\end{equation}
and

\begin{equation}
v_{1\;[\pm]}^{(\pm)}=\frac{2-\sqrt{-\Omega_\Lambda}\;v_2^{(\pm)}u_{1\;[%
\pm]}^{(\pm)}}{\sqrt{-\Omega_\Lambda}\;u_2^{(\pm)}},
\end{equation}
respectively.
Effective quintessence potential $\bar{W}(\phi)$ from field equations (2.4) or (2.5) and equations (2.7) and (2.13) can be written 
\begin{equation}
\bar{W}(\phi)=\bar{B}^2 e^{-\sigma\phi}-\bar V_0, 
\end{equation}
so the ansatz (2.18) establishes an interesting relationship between the value $\bar{V}(0)=\bar{B}^2$ of the exponential potential
\begin{equation}
\bar{V}(\phi)=\bar{B}^2 e^{-\sigma\phi}, 
\end{equation}
and the value $\bar{W}(\infty)=-\bar V_0$ towards which $\bar{W}(\phi)$ asymptotes. Other ansatze could have been taken, however, this one notably simplifies equations (for instance, eq.(2.17)). Also, as we will see later, many of the cosmological parameters result independent of $n$, avoiding fine tuning respect to this parameter. 

\section{Analysis of results}

Since $\sqrt{1+q_{0}-1.5\Omega _{m_{0}}}$ should be real (see equation
(2.21)) then, the following constrain on the present value of the
deceleration parameter follows

\begin{equation}
q_0\geq -1+1.5\Omega_{m_0}.
\end{equation}

\bigskip It can be noticed that the constants (and, consequently, the
solutions) depend on 4 physical parameters: $\Omega _{m_{0}}$, $\Omega
_{\Lambda }$, $q_{0}$ and on the positive real number $n$. 

After making a detailed study, it was determined that the only relevant cosmological magnitude that has a sensible dependence on parameter $n$ is the state parameter $\omega$. We used $\Omega_{m_{0}}=0.3$ and $q_0=-0.44$.
 
Figure 1 shows the evolution of the scale factor for $\Omega _{\Lambda
}=-0.15$. It can be shown both
algebraically and graphically that the evolution of the universe is independent of $n$, but not writen
for the sake of simplicity.
This results favour the formulation of both quantum field and string
theories. A breakdown of
 perturbative quantum field theory in
spacetimes with accelerated expansion is known to occur \cite{ssyy}.On
the other hand,
 an eternally accelerating universe seems to be at odds
with string theory, because of the impossibility of formulating the
S-matrix. In a deSitter space the presence of an event horizon, signifying
causally disconnected regions of space, implies the absence of asymptotic
particle states which are needed to define transition amplitudes. It is also interesting that Sen and Sethi, using an ansatz for the scale factor that produces future deceleration, obtain from field equations that the quintessence potential should be a double exponential plus a constant \cite{sese}. 
 
Figure 2 shows the behaviour of the deceleration parameter as function of the
redshift z for the
same values of the parameters. This figure shows an early stage of
deceleration and  a current epoch of acceleration. A transition from an accelerated phase
to a decelerated one is seen approximately for z =0.5. We appreciate an increase of
the deceleration parameter upon increasing the value of z. This points at a past epoch in the evolution when gravity of the dark energy was attractive. As follows from figure 1, aceleration is not eternal: in the future $q>0$ again,
which gives rise to the collapse.

Figure 3 shows the evolution of the state parameter of the effective quintessence field $\omega
_{\phi }$. It's noticeable that the effective quintessence field has state
parameter $\omega_{\phi }$ near $-1$ today, which means that its behaviour is similar to the ''pure'' cosmological constant, as a vacuum fluid. 
If we are to explain the
very desirable for today's cosmology recent and future deceleration obtained
in our model, it's important to look at the dynamical quintessence field. We
see that in the recent past $\omega_{\phi }>0$, which implies that
quintessence field behaved (or simply was) like ordinary atractive matter,
giving rise to the logical deceleration. In the future this will happen
again ($\omega_{\phi }>0)$ , with the consequent deceleration.

Now we proceed to analyze how our solution reproduces some experimental results.
With this purpose, in Fig. 4 we plot the distance modulus $\delta (z)$ vs
redshift z, calculated by us and the one obtained with the usual model with
a constant $\Lambda $ term. The relative deviations are of about 0.5$\%$.

So far, we have investigated one of the several possible branches of the solution, leaving for the future the investigation of the others.

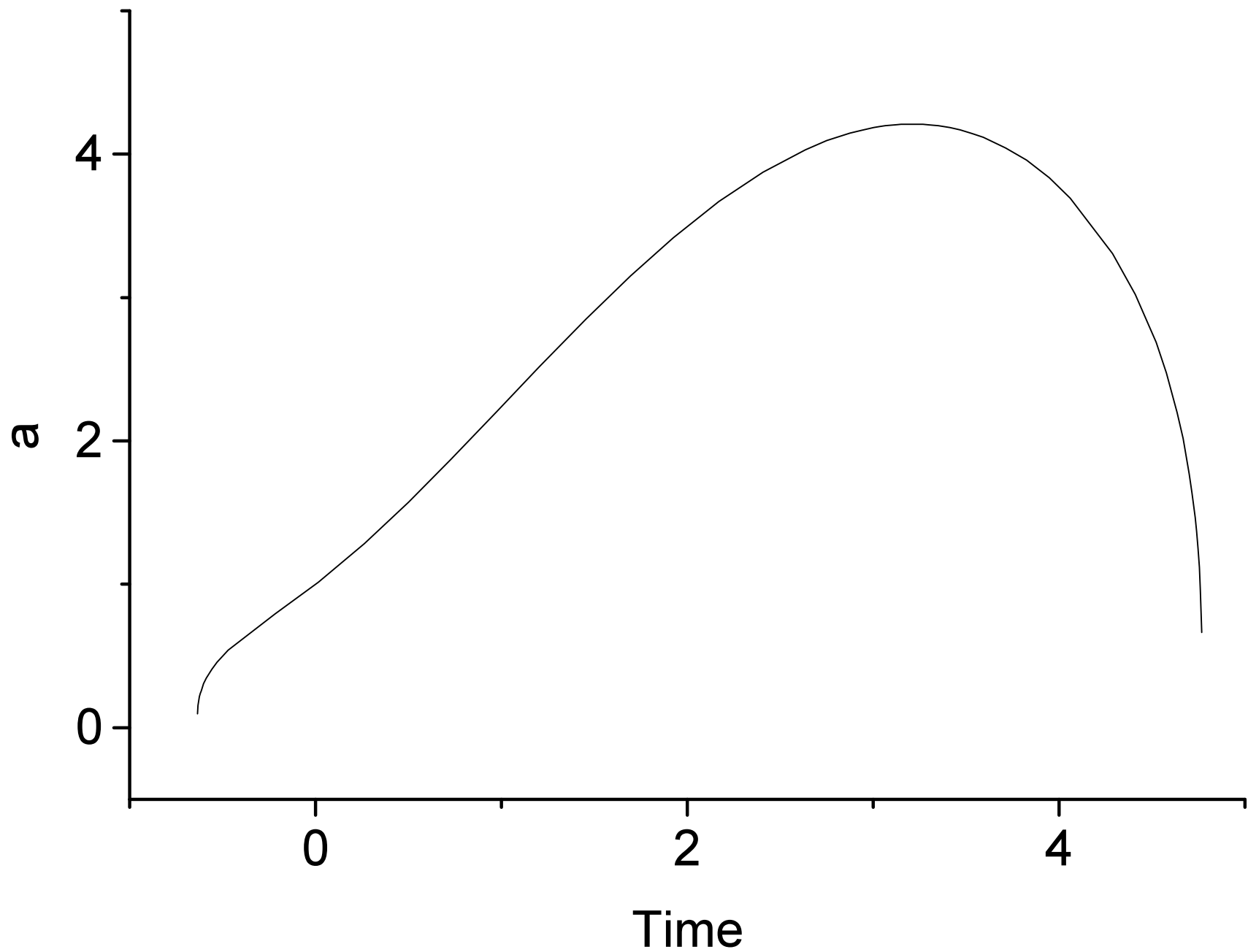
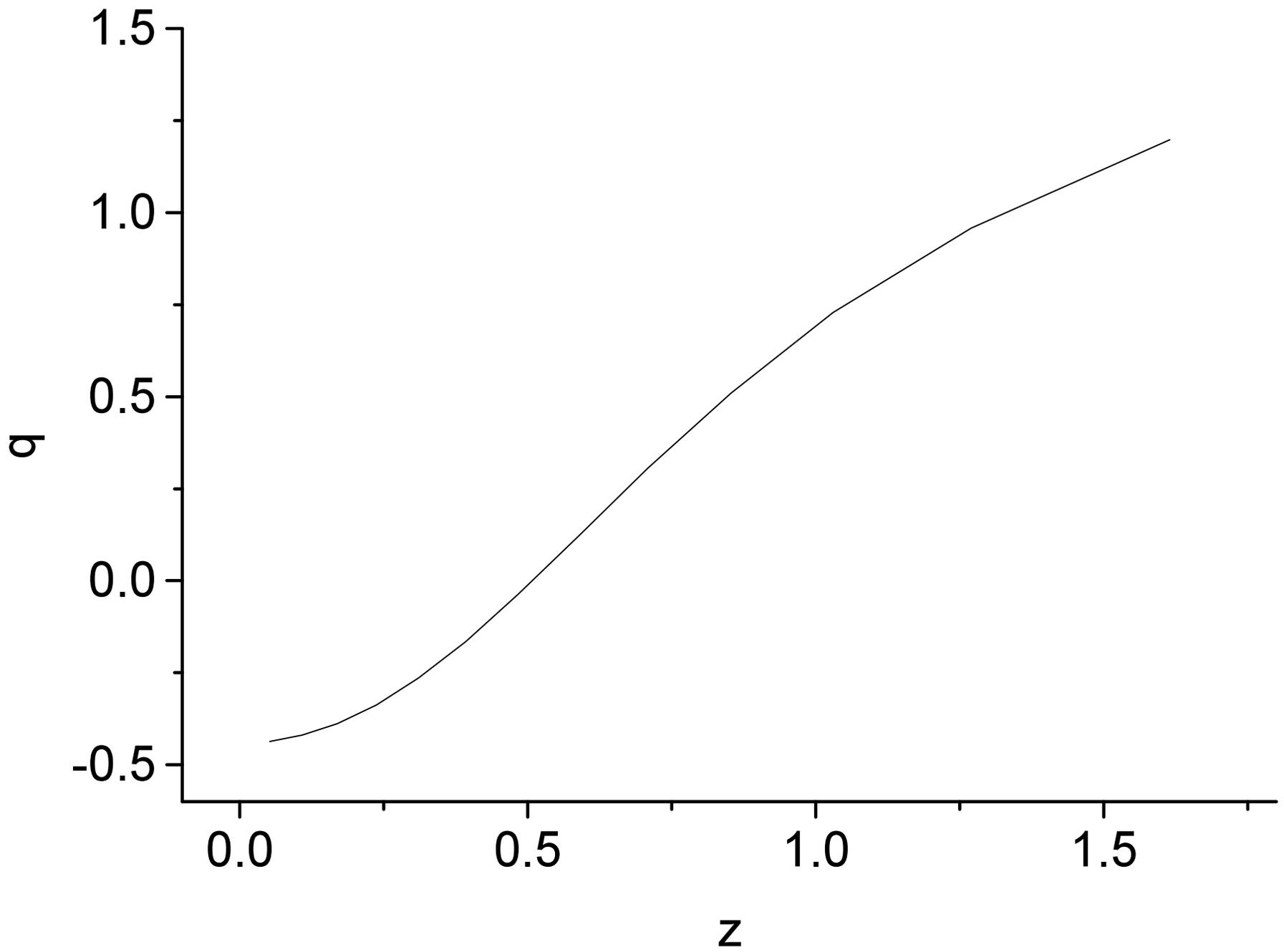
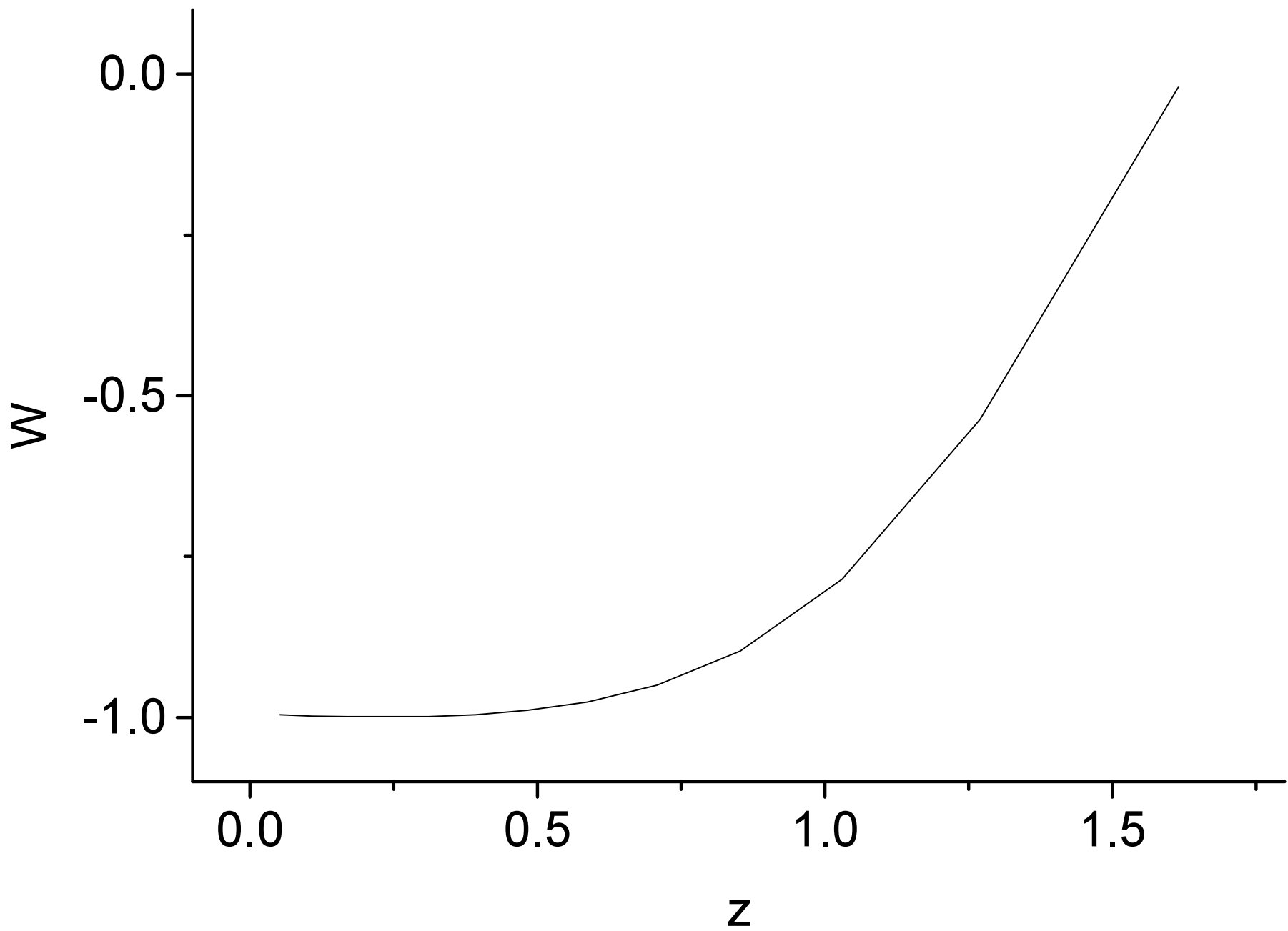
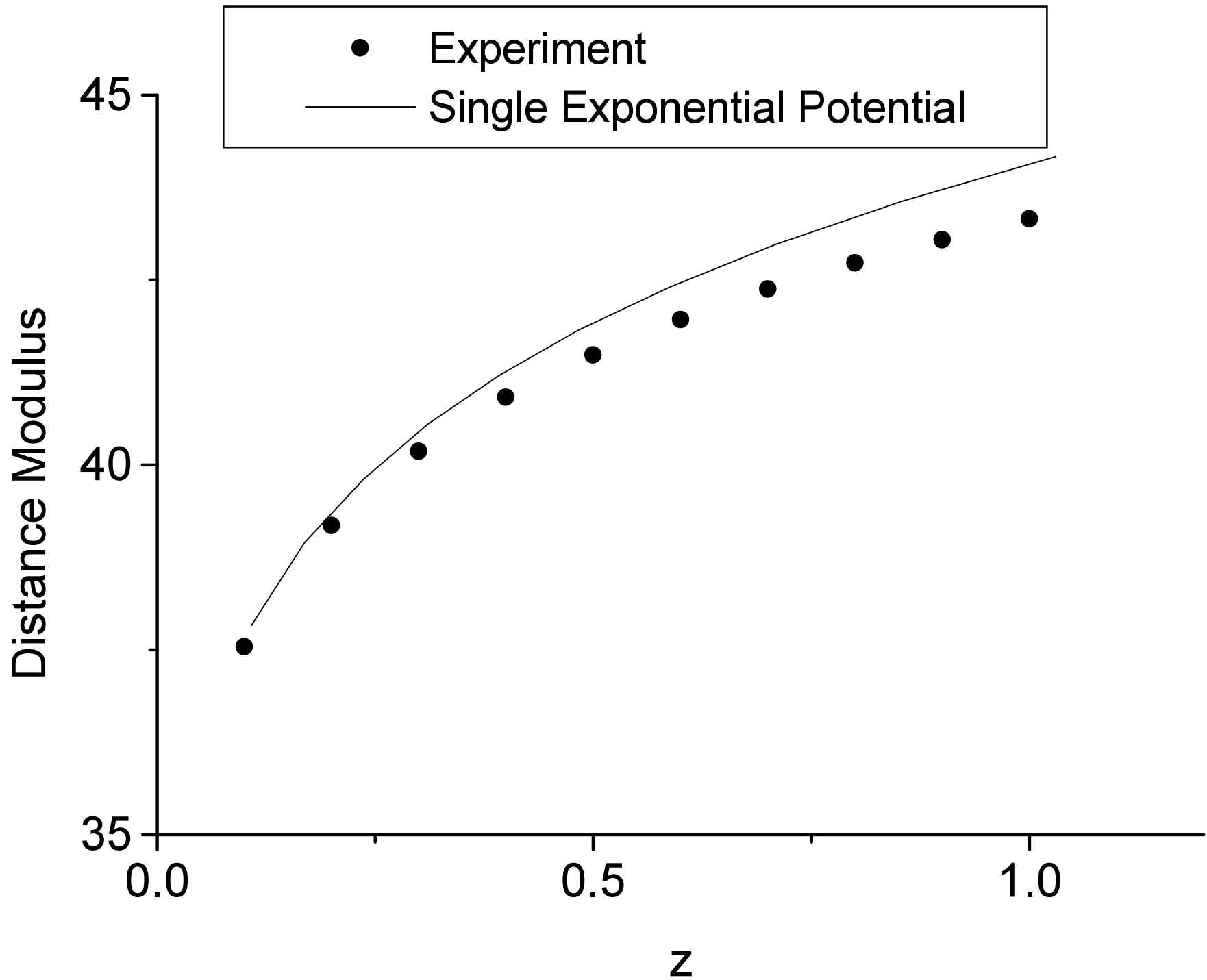
\end{document}